\documentclass[12pt]{article}
\usepackage{amssymb, amsmath}
\title{\bf Generalizations of Yang-Mills Theory with Nonlinear
Constitutive Equations}
\author{\bf  Gerald A. Goldin\footnote{e-mail: gagoldin@dimacs.rutgers.edu}
 \\  \it  Departments of Mathematics and Physics, Rutgers University, \\
 \it  Busch Campus, Piscataway, New Jersey 08854\\
 \bf  Vladimir M. Shtelen\footnote{e-mail: shtelen@math.rutgers.edu}\\
  \it   Department of Mathematics, Rutgers University, Busch Campus,\\
 \it   Piscataway, New Jersey 08854 }
 \date{}
 
\begin{document}
 \setcounter{section}{0}
 \renewcommand{\theequation}{\arabic{section}.\arabic{equation}}
 \maketitle
 \begin{abstract}{We generalize classical Yang-Mills
 theory by extending nonlinear constitutive equations for Maxwell
 fields to non-Abelian gauge groups. Such theories may or may not be
 Lagrangian. We obtain conditions on the constitutive equations specifying
 the Lagrangian case, of which recently-discussed non-Abelian
 Born-Infeld theories are particular examples. Some models in our class
 possess nontrivial Galilean ($c~\to~\infty$) limits; we determine
 when such limits exist, and obtain them explicitly.}
\end{abstract}
\section{Introduction }
 General equations for nonlinear, classical electromagnetic fields
 in media can be written beginning with Maxwell's equations for
 ${\bf E}$, ${\bf B}$, ${\bf D}$, and ${\bf H}$, and replacing the
 usual, {\it linear\/} constitutive equations by more general, {\it nonlinear\/}
 equations respecting Lorentz covariance. A general form for such
 systems was described by Fuschych, Shtelen, and Serov;
 familiar special cases include Born-Infeld
 and Euler-Kockel electrodynamics \cite{FSS,J99}. In earlier work we
 showed that certain nonlinear constitutive equations have well-defined Galilean-covariant
 limits as the
 speed of light $c \to \infty$, so that all four of Maxwell's equations remain valid \cite{GS1}.
 This is in sharp
 contrast to linear electrodynamics, where Maxwell's equations
 are well-known to be incompatible with Galilean relativity \cite{LB73,BH99}.
 Since classical Yang-Mills theory can be understood as an extension
 of classical electromagnetism to non-Abelian gauge potentials, it
 is natural to similarly extend Maxwell fields with nonlinear
 constitutive equations, and to ask whether such
 extensions also may have Galilean-covariant limits when $c \to \infty$.

Non-Abelian generalizations of the Born-Infeld Lagrangian \cite{BI}
(an excellent review of classical Born-Infeld theory is in \cite{BB})
have been known for some time, and recently have attracted renewed
interest \cite{H,Ts,P,KG,K}. In this paper we take a different
approach, deriving generalizations of classical Yang-Mills theory
 as non-Abelian extensions of Maxwell systems
 together with Lorentz-covariant, (in general) nonlinear constitutive equations.
 Standard Yang-Mills theory is a special case in this class of theories,
 with linear constitutive equations. Particular nonlinear constitutive
 equations correspond to the non-Abelian  Born-Infeld theories.
 Our approach has the important
 advantage that it is general enough to include Lagrangian and
 non-Lagrangian theories. In addition, since we directly
 generalize {\it nonlinear} Maxwell systems, we
 have the possibility of obtaining nontrivial
 Galilean-covariant (nonrelativistic) limits as $c \to \infty$.

 In Sec. $\!2$ we review Maxwell's equations for media,  and
characterize
 the family of nonlinear constitutive equations that result in  theories
 obtained from  invariant Lagrangians. In Sec. $\!3$ we generalize
 appropriately from $U(1)$ to non-Abelian gauge theory.
 In Sec. $\!4$ we consider the nonrelativistic $c \to
 \infty$ limit. Then we show that with necessary modifications,
 certain Born-Infeld (Abelian or non-Abelian)
 Lagrangian functions lead to nontrivial theories having such a limit.
 We state our conclusions in Sec. $\!5$.

\section{Nonlinear Electrodynamics }
\setcounter{equation}{0}
Here we use SI units, so that $c$ does not enter the definition of
${\bf E}$ or ${\bf B}$. We begin with
the usual metric tensor $g^{\mu\nu}={\textrm{diag}}\,(1,-1, -1, -1)$,
$x^{\mu}=(ct, {\bf x})$,
$x_{\mu}=g_{\mu\nu} x^{\nu}=(ct, -{\bf x})$, and
$x_{\mu}x^{\mu}=c^2t^2-{\bf x}^2$. We have $\partial_{\mu} \equiv
\partial/{\partial x^{\mu}}=[(1/c)\partial/{\partial
t},
 {\bf \nabla}]$, and we use the antisymmetric Levi-Civita
tensor $\varepsilon^{\alpha\beta\mu\nu}$ with
$\varepsilon^{0123}=1$.

The tensor fields constructed from vectors ${\bf E},\ {\bf B}, \
{\bf D}, \   {\bf H}$ are
 $$F_{\alpha\beta}=\left(\begin{array}{cccc}0&\frac{1}{c}E_1&\frac{1}{c}E_2&\frac{1}{c}E_3\\-
\frac{1}{c}E_1&0&-B_3&B_2\\
-\frac{1}{c}E_2&B_3&0&-B_1\\-\frac{1}{c}E_3&-B_2&B_1&0
  \end{array}\right), \quad
 F^{\alpha\beta}=g^{\alpha\mu}g^{\beta\nu}F_{\mu\nu}\,,$$

\begin{equation}
  {\mathcal{F}}^{\alpha\beta}=\frac{1}{2}\varepsilon^
 {\alpha\beta\mu\nu}F_{\mu\nu}= \left(\begin{array}{cccc}0&-B_1&-B_2&-B_3
 \\B_1&0&\frac{1}{c}E_3&-\frac{1}{c}E_2\\ B_2&-\frac{1}{c}E_3&0&\frac{1}{c}E_1\\
B_3&\frac{1}{c}E_2&-\frac{1}{c}E_1&0 \end{array}\right), \quad
{\mathcal{F}}_{\alpha\beta}=g_{\alpha\mu}g_{\beta\nu}{\mathcal{F}}^{\mu\nu}\,,
\nonumber
\end{equation}
\begin{equation}
G^{\alpha\beta}=
 \left(\begin{array}{cccc}0&-cD_1&-cD_2&-cD_3\\cD_1&0&-H_3&H_2\\
 cD_2&H_3&0&-H_1\\cD_3&-H_2&H_1&0
 \end{array}\right)\,,  \quad
 G_{\alpha\beta}=g_{\alpha\mu}g_{\beta\nu}G^{\mu\nu}\,.
 \label{matrices}
\end{equation}
 Maxwell's  equations for media in SI units take the form \cite{J99}
 \begin{equation}
 {\bf \nabla} \times {\bf E} =-\frac{\partial \bf B}{\partial t}, \quad
 {\bf \nabla} \cdot {\bf B} =0,\quad
 {\bf \nabla} \times {\bf H} =\frac{\partial \bf D}{\partial t} + {\bf j},
 \quad {\bf \nabla} \cdot {\bf D} =\rho, \label{me}
 \end{equation}
 or in covariant form
 \begin{equation}
 \partial_{\alpha}{\mathcal{F}}^{\alpha\beta}=0; \quad
 \partial_{\alpha}G^{\alpha\beta}=j^{\beta}, \quad
 \mathrm{where}\,\,
 j^{\beta}=(c\rho, {\bf j}). \label{ME}
 \end{equation}
The first equation in (\ref{ME}) allows us to introduce
$A_{\mu}=(\phi, -{\bf A})$ so that
 \begin{equation}
 {\mathcal{F}}^{\alpha\beta}=\epsilon^{\alpha\beta\mu\nu}
 \partial_{\mu}A_{\nu}, \quad F_{\mu\nu}=
 \partial_{\mu}A_{\nu}-\partial_{\nu}A_{\mu}.               \label{F}
 \end{equation}
 This system is invariant under the Lorentz
 group as well as the Galilei group;
 the choice between these symmetries
rests in the
 constitutive equations \cite{FSS,GS1,LB73}. The Lorentz invariant constitutive equations
 are
 \begin{equation}
 {\bf D} = M{\bf B} + \frac{1}{c^2}N{\bf E}, \quad {\bf H}=N{\bf B} - M{\bf
 E},\label{ceq}
 \end{equation}
 or in  covariant form
 \begin{equation}
 G^{\mu\nu}= NF^{\mu\nu}+ cM{\mathcal{F}}^{\mu\nu}
 \equiv M_1\frac{\partial I_1}{\partial F_{\mu\nu}}+ M_2\frac{\partial I_2}{\partial F_{\mu\nu}} ,\label{ce}
 \end{equation}
 where $M$ and $N$, or equivalently $M_1$ and $M_2$, are functions of the Lorentz invariants
\begin{equation}
I_1 ={\bf B}^2 -\frac{1}{c^2}{\bf E}^2=\frac{1}{2}~F_{\mu\nu}
F^{\mu\nu}, \quad I_2= {\bf B}\cdot{\bf E} =
-\frac{c}{4}~F_{\mu\nu}{\mathcal{F}}^{\mu\nu}.\label{LI}
\end{equation}
The standard Maxwell equations for the vacuum correspond to $M=0$,
$N\,= \,\mathrm{constant}\,=(\mu_0)^{-1}$, with
$c^2=(\mu_0\epsilon_0)^{-1}$.
A general form of
 an invariant Lagrangian for a nonlinear  theory given by Eqs. $\!$(\ref{ME})
 and (\ref{ce}) may be written
 ${\mathcal{L}}={\mathcal{L}}(I_1, \ I_2)$,
 where $I_1, \ I_2$ are given by (\ref{LI}). In such a theory, the tensor
 $G^{\mu\nu}$ becomes
\begin{equation}
 G^{\mu\nu}= -\frac{\partial{\mathcal{L}}}{\partial F_{\mu\nu}}=-\left(\frac{\partial{\mathcal{L}}}
 {\partial I_1}\right)\,2F^{\mu\nu}+c
 \left(\frac{\partial{\mathcal{L}}}{\partial I_2}\right)
 {\mathcal{F}}^{\mu\nu}\,.
 \label{lce}
 \end{equation}
 In the above, following e.g. Ref. \cite{BB}, the derivatives in (\ref{ce})
 are evaluated by first imposing in Eqs. (\ref{LI}) the constraints
 $F_{\mu\nu}=-F_{\nu\mu}$, $F^{\alpha\beta}=g^{\alpha\mu}g^{\beta\nu}F_{\mu\nu}$, and
${\mathcal{F}}^{\alpha\beta}=\frac{1}{2}\varepsilon^
 {\alpha\beta\mu\nu}F_{\mu\nu}$,
and then taking the partial derivatives of $I_1$ and $I_2$; thus
$$\frac{\partial{I_1}}{\partial F_{\mu\nu}}=2F^{\mu\nu}, \quad
 \frac{\partial{I_2}}{\partial F_{\mu\nu}}=-c\,{\mathcal{F}}^{\mu\nu}.$$
Comparison of Eqs. $\!$(\ref{lce}) with Eqs. $\!$(\ref{ce}) yields
the conditions
 \begin{equation}
 - 2\, \frac{\partial{\mathcal{L}}}{\partial I_1}=N, \quad
 \frac{\partial{\mathcal{L}}}{\partial I_2}=M,
 \end{equation}
 from which the compatibility condition for a Lagrangian theory
 reads,
 \begin{equation}
 2\, \frac{\partial M}{\partial I_{1}}+ \frac{\partial N}{\partial I_{2}}=0.
 \end{equation}
 The well-known Born-Infeld Lagrangian is usually written as
\begin{equation}
 {\mathcal{L}_{BI}}=\frac{b^2}{\mu_0 c^2}~(1-R), \quad R=
 \sqrt{1+\frac{c^2}{b^2}I_1-\frac{c^2}{b^4}I^2_2}\, ,
 \label{LBI}
 \end{equation}
where $b$ is a maximum electric field strength (in the absence of
magnetic field). If $\,b^2$ is very much larger than $\,{\bf
E}^2\,$ and $\,c^2{\bf B}^2$, then $\,\mathcal{L}_{BI} \approx
-(1/2\mu_0)I_1$ and we recover linear Maxwell theory. But
anticipating the discussion in Sec. $\!4$, we remark here that in
the limit as $\,c \to \infty$, $\mathcal{L}_{BI}$ tends to zero;
while $\,c\mathcal{L}_{BI}$ approaches a well-defined, non-zero
limit.

Another example is Euler-Kockel electrodynamics \cite{J99}.  Here,
in the first approximation, one has
 $M= 7\lambda (\mu_0)^{-1}I_2$ and $N=(\mu_0)^{-1}(1-2\lambda
 I_1)$,
  where $\lambda$ is a small parameter.
  The corresponding Lagrangian takes the form
  \begin{equation}
  {\mathcal{L}}=-\frac{1}{2\mu_0}I_1+2\lambda
  I_1^2+\frac{7\lambda}{2\mu_0}I^2_2\,,
  \end{equation}
  which, we remark, coincides
  with the ``toy model'' generalization of the Maxwell
  Lagrangian discussed by DeLorenci {\it et al \/}\cite{toy}.
\section{Generalization of Yang-Mills Theory}
 \setcounter{equation}{0}
To generalize the nonlinear electrodynamics described in Sec.
$\!\!2$ to non-Abelian gauge
 theory, we replace as usual the partial derivative $\partial_{\mu}$ by the
 commutator with the covariant derivative $D_{\mu}$; i.e.
 $\partial_{\mu} \to [D_{\mu}, \quad ]$, where
 \begin{equation}
D_{\mu}=\partial_{\mu} + igT^{\ell}W^{\ell}_{\mu}\,,
 \end{equation}
  $g$ is the YM coupling constant, the $T^{\ell}$ are the $N^2-1$ generators of
  $SU(N)$, and summation over $\ell$ is assumed. Then
  \begin{equation}
  [D_\mu, D_\nu] = igF_{\mu \nu}\,, \, \,\mathrm{where}\,\, F_{\mu \nu} =
  T^{\ell} F^{\ell}_{\mu \nu}\,.
  \end{equation}

The field equations  of the non-Abelian theory generalizing the nonlinear
Maxwell equations
 (\ref{ME}) and (\ref{ce}) take the form
 \begin{equation}\label{e}
[D_{\mu}, {\mathcal{F}}^{\mu\nu}]=0, \quad
 [D_{\mu},  G^{\mu\nu}]= J^{\nu}\,,
 \end{equation}
 where $J^{\nu}$ is an external current, and
 where the constitutive equations are to be written in a new way.
 Letting $u_s\,\, (s = 1,2,3, \dots , m)$  be a set of independent
 invariant functions of
the Yang-Mills fields, we write
 \begin{equation}
 G^{\ell \,\mu\nu}=\sum_{s=1}^{m}M_s(u_1,
u_2,..., u_m)~\frac{\partial u_s}{\partial F^{\ell}_{\mu\nu}}\,,
\label{b}
\end{equation}
where the $M_s$ are functions of the invariants. For the gauge
group $SU(N)$ we have no fewer than $m=5N^2-11$ independent
invariants, using the following simple argument of Roskies:  Since
the gauge group $SU(N)$ has $N^2-1$ parameters, and the Lorentz
group has $6$ parameters, the number of components of $
F^{\ell}_{\mu\nu}$ is $6(N^2-1)$. One can choose a Lorentz frame
and an $O(N)$ frame in which $6+(N^2-1)= 5+N^2$  components
vanish. There will then be $6(N^2-1)-(5+N^2)=5N^2-11$ remaining
components. Any invariant could be evaluated in this special
frame, and therefore could be a function of these $5N^2-11$
components. In particular, there are $9$ independent invariants
for $SU(2)$ \cite{R77}:
\[
u_1=\textrm{tr}(K), \quad
u_2=-\frac{1}{2}\textrm{tr}(J),
\]
$$u_3=\frac{1}{4}\textrm{tr}(J^2), \ u_4=-\det(J), \  u_5=
\textrm{tr}(K^2), \ u_6=\det(K), \ u_7=\textrm{tr}(JK),$$
\begin{equation}
u_8=\frac{1}{6}\varepsilon_{ijk}F_{\mu}^{i~\nu}F_{\nu}^{j~\rho}F_{\rho}^{k~\mu},
\quad u_9=-\frac{c}{6}\varepsilon_{ijk}{\mathcal{F}}_{\mu}^{i~\nu}
{\mathcal{F}}_{\nu}^{j~\rho}{\mathcal{F}}_{\rho}^{k~\mu},
\label{ymi}
\end{equation}
where
\[
K_{ij}=\frac{1}{2}~F_{\mu\nu}^{i}F^{j~\mu\nu}={\bf B}^i\cdot{\bf B}^j-\frac{1}{c^2}{\bf E}^i\cdot{\bf E}^j,
\]
\begin{equation}
J_{ij}=\frac{c}{2}~F_{\mu\nu}^{i}{\mathcal{F}}^{j~\mu\nu}= -[{\bf
B}^i\cdot{\bf E}^j+{\bf B}^j\cdot{\bf E}^i]\,, \label{JK}
\end{equation}
with $i,j,k = 1,2,3$ being the $SU(2)$ algebra indices. The factors of $c$ in Eqs.
$\!$(\ref{ymi})-(\ref{JK}) have been introduced so that in all
cases, the limit $c \to \infty$ results in survival of the leading
terms.

With a little help from Maple, we calculated the explicit form of
the invariants $u_1,..., u_9$ in Eqs. $\!$(\ref{ymi}). In the
notation that follows, the ${{\bf B}^\ell}$
 (gauge components $\ell=1,2,3$) are vectors with
 spatial components $B^{\ell}_1,~B^{\ell}_2$, and $B^{\ell}_3$.
Here are these Lorentz YM gauge invariants:
\begin{equation}
u_1=\sum_{\ell \,= 1}^3 (\,{\bf B}^{\ell}\cdot {\bf B}^{\ell}
-\frac{1}{c^2}{\bf E}^{\ell}\cdot {\bf E}^{\ell}\,)\,, \nonumber
\end{equation}
\begin{equation}
u_2=  \sum_{\ell \,= 1}^3 {\bf B}^{\ell}\cdot{\bf E}^{\ell}\,,
\nonumber
\end{equation}
\[
u_3\,=\,({{\bf B}^1}\cdot{{\bf E}^1})^2 \,+\, ({{\bf
B}^2}\cdot{{\bf E}^2})^2\,+\,({{\bf B}^3}\cdot{{\bf E}^3})^2
\]
\begin{equation}
+\,\frac{1}{2}\,[\,({{\bf B}^1}\cdot{{\bf E}^2}+{{\bf
B}^2}\cdot{{\bf
 E}^1})^2\,+\,
 ({{\bf B}^2}\cdot{{\bf E}^3}+{{\bf B}^3}\cdot{{\bf
 E}^2})^2\,+\,
  ({{\bf B}^3}\cdot{{\bf E}^1}+{{\bf B}^1}\cdot{{\bf
  E}^3})^2\,]\,, \nonumber
\end{equation}
\begin{equation}
u_4=\det(J), \ {\textrm{where}} \ J_{ij}=-[{\bf B}^i\cdot{\bf
E}^j+{\bf B}^j\cdot{\bf E}^i], \nonumber
\end{equation}
\[
u_5\,=\,({{\bf B}^1}\cdot{{\bf B}^1}-\frac{1}{c^2}{\bf E}^{1}\cdot
{\bf E}^{1})^2 + ({{\bf B}^2}\cdot{{\bf B}^2}-\frac{1}{c^2}{\bf
E}^{2}\cdot {\bf E}^{2})^2 + ({{\bf B}^3}\cdot{{\bf
B}^3}-\frac{1}{c^2}{\bf E}^{3}\cdot {\bf E}^{3})^2
\]
\begin{equation}
+\,2\,[\,({{\bf B}^1}\cdot{{\bf B}^2}-\frac{1}{c^2}{{\bf
E}^1}\cdot{{\bf E}^2})^2+
 ({{\bf B}^1}\cdot{{\bf B}^3}-\frac{1}{c^2}{{\bf E}^1}\cdot{{\bf E}^3})^2+
 ({{\bf B}^2}\cdot{{\bf B}^3}-\frac{1}{c^2}{{\bf E}^2}\cdot{{\bf E}^3})^2] \,,
 \nonumber
\end{equation}
\begin{equation}
u_6=\det(K), \ {\textrm{where}} \ K_{ij}= {\bf B}^i\cdot{\bf
B}^j-\frac{1}{c^2}{\bf E}^i\cdot{\bf E}^j, \nonumber
\end{equation}
$$u_7=({{\bf B}^1}\cdot{{\bf B}^2}-\frac{1}{c^2}{\bf E}^{1}\cdot {\bf E}^{2})(
{\bf E}^{1}\cdot {\bf B}^{2}+{\bf B}^{1}\cdot {\bf E}^{2})+
({{\bf B}^1}\cdot{{\bf B}^3}-\frac{1}{c^2}{\bf E}^{1}\cdot {\bf E}^{3})(
{\bf E}^{1}\cdot {\bf B}^{3}+{\bf B}^{1}\cdot {\bf E}^{3})$$
\begin{equation}
+\, ({{\bf B}^2}\cdot{{\bf B}^3}-\frac{1}{c^2}{\bf E}^{2}\cdot
{\bf E}^{3})( {\bf E}^{2}\cdot {\bf B}^{3}+{\bf B}^{2}\cdot {\bf
E}^{3}) \,, \nonumber
\end{equation}
\begin{equation}
 u_8= ({\bf B}^1 \times {\bf B}^2)\cdot {\bf B}^3 -
 \frac{1}{c^2}[({\bf E}^1 \times {\bf E}^2)\cdot {\bf B}^3+
({\bf E}^2 \times {\bf E}^3)\cdot {\bf B}^1 +
 ({\bf E}^3 \times {\bf E}^1)\cdot {\bf B}^2], \nonumber
\end{equation}
\begin{equation}
 u_9= - \frac{1}{2}\,\epsilon_{ijk}({\bf B}^i \times {\bf
B}^j)\cdot {\bf E}^k + \frac{1}{c^2}({\bf E}^1 \times {\bf E}^2)\cdot {\bf E}^3\,.
\end{equation}
Note that $u_2, u_3,$ and $u_4$  are independent of $c$, and therefore they will be the same
in the Galilean limit ($c\to\infty$).

There are now Lagrangian and non-Lagrangian theories
determined by Eqs. $\!$(\ref{e})-(\ref{b}). In a Lagrangian theory
the constitutive equations are
\begin{equation}
G^{\ell\,\mu\nu}= -\frac{\partial{\mathcal{L}}}{\partial
F_{\mu\nu}^{\ell}}=
-\sum_{s=1}^{m}\frac{\partial{\mathcal{L}}}{\partial
u_s}\frac{\partial u_s}{\partial F_{\mu\nu}^{\ell}}\,.
\label{nace}
\end{equation}
Thus Eqs. $\!$(\ref{e})-(\ref{b}) determine a Lagrangian theory if
and only if the coefficients $M_s$ in (\ref{b}) can be written as
$ M_s=-{\partial{\mathcal{L}}}/{\partial u_s} $ for some
scalar-valued function ${\mathcal{L}}={\mathcal{L}}(u_1, u_2,...,
u_m)$. The corresponding restrictions on the $M_s$ are the
compatibility conditions resulting from the equalities of the
mixed derivatives of ${\mathcal{L}}$ with respect to $u_r$ and $\
u_s$; i.e., $\partial M_s/\partial u_r = \partial M_r /
\partial u_s$ ($\forall \ r,s = 1,2,\ldots,m)$.

In particular, one obtains non-Abelian versions of Born-Infeld or
Euler-Kockel theory by taking various generalizations of the
respective Lagrangians discussed in Sec. $\!2$. For example, a
Born-Infeld Lagrangian proposed in \cite{H} for non-Abelian
chromodynamics (CD) is given by
\begin{equation}
{\mathcal{L}_{BICD}}=\frac{b^2}{\mu_0 c^2}~(1-R_{CD}), \quad
R_{CD}=
 \sqrt{1+\frac{c^2}{b^2}u_1-\frac{c^2}{3b^4}(u^2_2+2u_3)}\,\,.
 \label{LBICD}
\end{equation}
In the Abelian case $\,u_1\,$ is $\,I_1$, $\,u_2\,$ is $I_2$, and
$\,u_3\,$ reduces to $\,u^2_2\,= \,I_2^2$; so that Eq.
\!(\ref{LBICD}) becomes the same as Eq. \!(\ref{LBI}).
\section{A Framework for Non-Abelian Galilean Theories}
\setcounter{equation}{0}
Let us consider  the nonrelativistic limit of the equations derived in Sec. 2.
Galilean symmetry transformations (the $c \to \infty$ limit
of Lorentz transformations) have the form
 \begin{equation}
 t'=t, \ {\bf x}'={\bf x}-{\bf v}t,\quad
 {({\bf E}^{\ell})'= {\bf E}^{\ell} + {\bf v}\times{\bf B}^{\ell}, \quad
 ({\bf B}^{\ell}})'= {\bf B}^{\ell}.\label{GT}
 \end{equation}
  As  is well known, there is no
 nonrelativistic limit of the
 standard Yang-Mills equations. This is because
 the linear constitutive equations
  $G^{\ell \,\mu\nu}~=~(1/\mu_0)~F^{\ell \,\mu\nu}$
  break  the Galilean symmetry.
 But our equations  (\ref{e})-(\ref{b}) {\it can\/} have
 a $c \to \infty$ limit,  provided the constitutive equations are
also nonlinear. One obtains such a Galilean non-Abelian gauge
theory from Eqs. $\!$(\ref{e})-(\ref{b}), writing these equations
explicitly in terms of ${\bf E}^{\ell},\  {\bf B}^{\ell}, \ {\bf
D}^{\ell}$ and ${\bf H}^{\ell}$, and then taking the limit as $c
\to \infty$. The equations of motion (\ref{e}) will always be the
same as in the relativistic theory, as the factors of $c$ cancel;
only the constitutive equations (\ref{b}) will be different.

Here are  the Galilean YM gauge invariants $\hat u_1,\ldots \hat u_9$:
$$
\hat u_1= \sum_{\ell\,=1}^3 \,{{\bf B}^\ell}\cdot{{\bf B}^\ell}\,,
$$
\begin{equation}
\hat u_2=u_2= \sum_{\ell\,=1}^3 \,{{\bf B}^\ell}\cdot{{\bf
E}^\ell}\,, \nonumber
\end{equation}
\[
\hat u_3\,=\,u_3\,=\,({{\bf B}^1}\cdot{{\bf E}^1})^2 \,+\, ({{\bf
B}^2}\cdot{{\bf E}^2})^2\,+\,({{\bf B}^3}\cdot{{\bf E}^3})^2
\]
\begin{equation}
+\,\frac{1}{2}\,[\,({{\bf B}^1}\cdot{{\bf E}^2}+{{\bf
B}^2}\cdot{{\bf
 E}^1})^2\,+\,
 ({{\bf B}^2}\cdot{{\bf E}^3}+{{\bf B}^3}\cdot{{\bf
 E}^2})^2\,+\,
  ({{\bf B}^3}\cdot{{\bf E}^1}+{{\bf B}^1}\cdot{{\bf
  E}^3})^2\,]\,,\nonumber
\end{equation}
$$
\hat u_4=u_4=\det({\bf B}^i\cdot{\bf E}^j+{\bf B}^j\cdot{\bf E}^i),
$$
$$
\hat u_5\,=\,({{\bf B}^1}\cdot{{\bf B}^1})^2+
({{\bf B}^2}\cdot{{\bf B}^2})^2 +
({{\bf B}^3}\cdot{{\bf B}^3})^2
  +2\,[\,({{\bf B}^1}\cdot{{\bf B}^2})^2+
 ({{\bf B}^1}\cdot{{\bf B}^3})^2+
 ({{\bf B}^2}\cdot{{\bf B}^3})^2],
$$
$$
\hat u_6=\det({\bf B}^i\cdot{\bf B}^j),
$$
$$\hat u_7=({{\bf B}^1}\cdot{{\bf B}^2})({\bf E}^{1}\cdot {\bf B}^{2}+{\bf B}^{1}\cdot {\bf E}^{2})+
({{\bf B}^1}\cdot{{\bf B}^3})({\bf E}^{1}\cdot {\bf B}^{3}+{\bf B}^{1}\cdot {\bf E}^{3})
({{\bf B}^2}\cdot{{\bf B}^3})(
{\bf E}^{2}\cdot {\bf B}^{3}+{\bf B}^{2}\cdot {\bf E}^{3}),
$$
$$\hat u_8= \frac{1}{6} \,\epsilon_{ijk}({\bf B}^i \times {\bf
B}^j)\cdot {\bf B}^k = ({\bf B}^1 \times {\bf B}^2)\cdot {\bf
B}^3\,, $$
\begin{equation}
\hat u_9= - \frac{1}{2}\,\epsilon_{ijk}({\bf B}^i \times {\bf
B}^j)\cdot {\bf E}^k\,. \label{hatu}
\end{equation}
Using (\ref{GT}) one can check directly (Maple helps)
that, indeed,
$\hat u_1,\ldots \hat u_9$ are Galilean invariants.

Let us look at some Born-Infeld theories in the Galilean limit. In
the Abelian case we obtain constitutive equations of the form of
Eq. \!(\ref{ceq}), with
\begin{equation}
M = \frac{I_2}{\mu_0 b^2R},\quad N = \frac{1}{\mu_0R}\,.
\end{equation}
In the limit as $\,c \to \infty$, we have $I_1 \to \hat{I}_1 =
\mathbf{B}^2$, and $I_2 = \hat{I}_2 = \mathbf{B}\cdot\mathbf{E}$.
But in this limit $\,R\,\approx (c/b) [\hat{I}_1 -
\hat{I}_2^2/b^2]^{1/2}$, so that $M$ and $N$ do not approach
well-defined nonzero limits. This suggests modification of the
Born-Infeld Lagrangian. For example, one possibility is to replace
$R$ in Eq. \!(\ref{LBI}) by
\begin{equation}
\tilde{R} = \sqrt{1+\frac{c^2}{b^2}\,[\,(1 + \lambda_1 c^2)
I_1-\frac{1}{b^2}(1+\lambda_2 c^2) I_2^{\,2}\,]\,}\,,
\end{equation}
where $\lambda_1$, $\lambda_2$ have the dimensionality of inverse
velocity squared. Then in the limit when $\,c \to \infty$, we
obtain the Galilean constitutive equations

\begin{equation}
 {\bf D} = \hat M{\bf B}, \quad {\bf H}=\hat{N}{\bf B} - \hat{M}{\bf
 E}\,,\label{nrelceq}
 \end{equation}
where
\begin{equation}
\hat{M} = \frac{ \lambda_2 \hat{I}_2}{\mu_0 b
\sqrt{\lambda_1\hat{I}_1 - \lambda_2\hat{I}_2^{\,2}/b^2}}\,,\quad
\hat{N} = \frac{b \lambda_1}{\mu_0 \sqrt{\lambda_1\hat{I}_1 -
\lambda_2\hat{I}_2^{\,2}/b^2}}\,.
\end{equation}

Similarly, in the non-Abelian case, we obtain a well-defined
Galilean limit for the Yang-Mills constitutive equations
(\ref{nace}) if we modify $R_{CD}$ in Eq. \!(\ref{LBICD}) to be
\begin{equation}
\tilde{R}_{CD} = \sqrt{1+\frac{c^2}{b^2}(1 + \lambda_1
c^2)u_1-\frac{c^2}{3b^4}(1 + \lambda_2 c^2)(u^2_2+2u_3)}\,\,.
\end{equation}
Then, with $\,c \to \infty$,
\begin{equation}
{\bf D}^\ell = \frac{ \lambda_2 (\hat{u}_2+2{\bf B}^\ell\cdot {\bf
E}^\ell)}{3\mu_0 b \sqrt{\lambda_1\hat{u}_1 -
\frac{\lambda_2}{3b^2}(\hat{u}_2^{\,2}+2\hat{u}_3)}}\,{\bf
B}^\ell\, \label{nace1}
\end{equation}
and
\begin{equation} {\bf H}^\ell=\frac{b \lambda_1}{\mu_0 \sqrt{\lambda_1\hat{u}_1 -
\frac{\lambda_2}{3b^2}(\hat{u}_2^{\,2}+2\hat{u}_3)}}\,{\bf B}^\ell
- \frac{ \lambda_2 (\hat{u}_2+2{\bf B}^\ell\cdot {\bf
E}^\ell)}{3\mu_0 b \sqrt{\lambda_1\hat{u}_1 -
\frac{\lambda_2}{3b^2}(\hat{u}_2^{\,2}+2\hat{u}_3)}}\,{\bf
 E}^\ell\,. \label{nace2}
 \end{equation}
 Note that in Eqs. \!(\ref{nace1})-(\ref{nace2}) there is no summation over
 $\ell$, while $\hat{u}_1, \hat{u}_2, \hat{u}_3$ are the Galilean invariants given by Eqs.
 \!(\ref{hatu}).

 We close this section with the remark that the nonlinear
gauge theory described here can be set up usefully with a
``Galilei friendly" metric tensor, respecting  the fact that space
and time require different units (independent of $c$) if the
Galilean limit is to be meaningful. By setting $\,\hat
g^{\mu\nu}={\textrm{diag}}
 (1/{c^2}, -1, -1, -1)$ and $\hat g_{\mu\nu}={\textrm{diag}}(c^2, -1, -1, -1)$,
so that  $\,x^{\mu}=(t, {\bf x}), \, x_{\mu}=\hat g_{\mu\nu}
x^{\nu}=(c^2t, -{\bf x})$, and $\, x_{\mu}x^{\mu}=c^2t^2-{\bf
x}^2$, we obtain in place of Eqs. (\ref{matrices}) matrix
expressions for $\,\hat{F}_{\alpha\beta},\,\hat
{\mathcal{F}}^{\alpha\beta},\,\hat G^{\alpha\beta},$ and $\hat
{\mathcal{G}}_{\alpha\beta} $ in terms of the fields $\,{\bf E}\,$
and $\,{\bf B}\,$ that involve no factors of $c$; while factors of
$\,1/c^2\,$ or $\,c^2\,$ occur in the expressions for the other
field strengths. With such a choice, taking the limit $c\to\infty$
in the relativistic equations is straightforward. The equations of
motion (\ref{e}) do not involve $c$ and do not change, while the
 constitutive equations (\ref{b}) change as $c\to\infty$.


\section{Conclusion}

 We have seen how it is possible to generalize nonlinear Maxwell
 systems directly to the case of non-Abelian gauge groups,
 thus obtaining generalized Yang-Mills theories associated with
 nonlinear constitutive equations for the fields. Such a theory
 may or may not be derivable from a Lagrangian function.
 Our construction allows for either situation, and permits one
 to determine directly from the constitutive
 equations whether or not a Lagrangian formulation is possible.

 In particular,
 our approach highlights the possibility of obtaining nontrivial
 Galilean-covariant (nonrelativistic) limits of these theories
 as $c \to \infty$. We have seen that
 such limits exist in some cases, but not in all.
 We believe they have potential application in contexts
 where Galilean theories are coupled with nonlinear
 electromagnetic fields or their non-Abelian counterparts---for
 example, in nonlinear Schr\"odinger theory as described in \cite{DGN1999}
 and discussed in \cite{GS1}, or in non-Abelian fluid mechanics
 \cite{Jackiw2002, BJLNP2002, GrundHar2004}. They are also
 potentially applicable, as noted in \cite{GS1}, to electromagnetic fields in
 condensed matter where the nonlinearity is extremely strong,
 and as effective, low-energy limits in string theory.

\section*{Acknowledgment}
GG would like to thank R. Kerner for interesting discussions and helpful
suggestions.

\end{document}